# Program Spectra Analysis in Embedded Software: A Case Study

Rui Abreu, Peter Zoeteweij and Arjan JC van Gemund

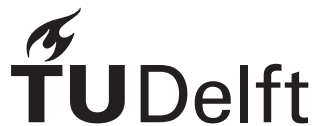

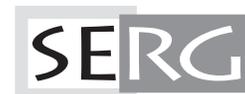

# Program Spectra Analysis in Embedded Software: A Case Study*

**Rui Abreu**      **Peter Zoeteweij**      **Arjan J.C. van Gemund**

Software Technology Department,
Faculty of Electrical Engineering, Mathematics, and Computer Science,
Delft University of Technology,
P.O. Box 5031, 2600 GA Delft, The Netherlands
{r.abreu, p.zoeteweij, a.j.c.vangemund}@ewi.tudelft.nl

## Abstract

Because of constraints imposed by the market, embedded software in consumer electronics is almost inevitably shipped with faults and the goal is just to reduce the inherent unreliability to an acceptable level before a product has to be released. Automatic fault diagnosis is a valuable tool to capture software faults without extra effort spent on testing. Apart from a debugging aid at design and integration time, fault diagnosis can help analyzing problems during operation, which allows for more accurate system recovery. In this paper we discuss perspectives and limitations for applying a particular fault diagnosis technique, namely the analysis of program spectra, in the area of embedded software in consumer electronics devices. We illustrate these by our first experience with a test case from industry.

## 1 Introduction

Modern consumer electronic systems such as household electronics, radios, MP3–players, mobile phones, etc., increasingly rely on embedded software content. Software-intensive products contain many residual faults due to their size, complexity and high pressure on development time. In the past almost all functionality of consumer electronics devices was implemented in hardware, but nowadays, new features are more and more implemented in software, and embedded software in consumer electronics is growing rapidly. To illustrate, a typical up-market television set has already 16MB of software inside. This is due the fact that nowadays televisions have much more functionalities than ten years ago, such as extensive video processing, interoperability with other systems and support for several video formats. There is also a great diversity in product features, user control, supported broadcasting standards and hardware technology. Increasingly shorter time-to-market is one fact in consumer electronics context and consequently, in order to accomplish deadlines, consumer electronics embedded software is usually shipped with several bugs. It can happen that these bugs are visible to the consumers.

Debugging software is very time consuming, and generally exhaustive testing in terms of all possible input combinations is simply impossible. Automatic fault diagnosis techniques may help developers to detect and find bugs faster. Furthermore, we believe that diagnosis techniques can also be applicable at run-time helping automatic recovery strategies. This paper presents our very first experiment with fault diagnosis in the context of a particular product line of television sets.

A prerequisite for diagnosing faults in a program is the acquisition of data about the system activity. Examples of possible data that would help to diagnose a program are traces of events such as function calls, execution of basic blocks (a set of statements) or even individual statements. The large volumes of data that are thus collected for realistic systems are processed further off-line to obtain a meaningful analysis of the system's behavior.

In embedded systems in general and particularly in the TV case, the possibilities of tracing are limited by the amount of available memory and by the bandwidth of the communication to extract the data from the TV. The possibilities for on-the-fly (online) data processing are further limited by the amount of idle CPU cycles. In this paper we therefore use *program spectra*, which are compact rep-resentations of full traces. Collecting program spectra typically costs little CPU time and generates small amounts of data. Further, we show how we managed to generate program spectra from the system and how we removed them from the TV without interfering with the normal behavior of the system.

---

*This work has been carried out as part of the TRADER project [1] under the responsibility of the Embedded Systems Institute. This project is partially supported by the Netherlands Ministry of Economic Affairs under the BSIK03021 program.

This paper is organized as follows. Section 2 introduces fault diagnosis concepts and definitions and related work. In this section we also present our approach to diagnose embedded software in consumer electronics. In Section 3 we introduce FRONT, the tool used to automatically instrument source, presenting one simple example to show its usage. Section 4 provides a case study describing one experiment carried out in the TV. In Section 5 we draw our conclusions from this very first experiment and we also give insights about future directions.

## 2  Fault Diagnosis Using Spectra

This section introduces the basic concepts of error detection and fault diagnosis, the principles of our experiments, and the related work in the error detection and diagnosis area.

### 2.1  Concepts & Definitions

Following Avizienis *et al* [2] we define a *failure* as an event that occurs when a delivered service deviates from correct service. A correct service is delivered when the service implements the system function. A system may fail because (1) it does not comply with the specification or (2) the specification did not adequately describe the required system function. A *fault* is the cause of an error in the system and an *error* is a system state that may cause a system failure.

To better understand these concepts consider the following example:

```
[S01]  a = 4
       ...
[S14]  if (x) b = a − 4;
       ...
[S48]  c = a / b;
```

The above piece of code will result in an error when x is true, due to the division by zero. If so, the failure will occur in S48, where the actual division takes place. This division by zero occurs because b has got the value zero. The statement that causes b to be zero is said to be the fault. Thus the fault may reside in S1 and/or S14.

A prerequisite to diagnose the root cause of an error is to have the set of events that caused the error. This set of events can be used then to reason about the system and eventually may allow us to find the exact root cause of the error. This process is usually called error detection and fault diagnosis.

In order to detect an error, we need therefore to introduce "sensors" in our system (for instance by means of source code instrumentation). These sensors can be for instance functions to trace the execution of statements, basic blocks or components. A *transaction* is a finite sequence of events and the information that comes from the sensors form a trace. A transaction either ends in an error ("fail") or not ("pass"). This information comes from the error detection mechanism and is to be used as input to a fault diagnosis algorithm.

Formally, diagnosis is the process of finding differences of the real outcome of the system and the expected outcome. Ideally the way to find the root cause of an error would be using a model of the system, i.e., comparing the behavior of the system against the specifications of the system (a different outcome could signal an error). However, software is not easy to model and its model could well be as complex as the system itself. Because of the complexity of software models, fault localization is usually implemented using a black-box approach instead of a white-box approach, i.e., the software components are analyzed as "black-box" in terms of their behavior to a given input.

To illustrate the concepts of error detection and fault diagnosis, consider Figure 1. $C_1$, $C_2$, $C_3$ and $C_4$ are software basic-blocks. Using a black-box approach, we can only measure input and output values of the basic-blocks.

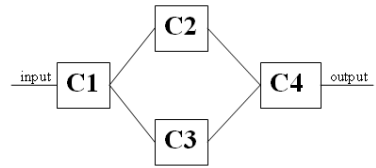

**Figure 1. Control-flow graph Basic-blocks**

If we couple a trace and error detection mechanism to the system represented by Figure 1, we would be able to trace the executed basic-blocks. The trace is labeled as passed ($r_\checkmark$) or failed ($r_\times$) using the error detection mechanism. Let $T_i$ denote trace (of transaction) $i$. If we were to collect the two following traces,

$$T_1 = C_1\ C_2\ C_4, r_\checkmark$$

$$T_2 = C_1\ C_3\ C_4, r_\times$$

Applying a simple diagnosis algorithm we would say the fault is probably in $C_3$.[1]

### 2.2  Program Spectra

In the above example, fault diagnosis involves a comparison of two traces of the system. Intuitively, traces are order-preserving accounts of events that we are interested in, possibly augmented with accurate timing information, for example, traces of function calls, or traces of branching decisions. Because the order of the events is preserved, and the number of events is typically large, traces are impractical for many applications.

---

[1] As the second trace may involve other data arguments, $C_1$ and $C_4$ are still suspect, although less probable.

A program spectrum is a compact representation, or projection, of a full trace of what happens during the execution of a program. While a trace can be seen as a specific kind of program spectrum (called the Execution Trace Spectrum in [3]), usually the order of the events is lost in the projection by accumulating the information for different occurrences of the same event. Examples of program spectra that have been reported in the literature are path spectra, counting the number of times each possible intra-procedural path is taken, and branch spectra, counting the number of times each possible branching decision is made. Execution profiles, as generated by tools as `gprof` and `gcov`, can also be seen as program spectra.

Just like in the example of Section 2.1, the basic idea of using program spectra for fault diagnosis is to analyze the differences between spectra for runs of a program in which an error occurred, and spectra for runs of a program in which no error occurred. Even though the ordering information is lost, comparing program spectra may still correlate certain events with the occurrence of an error, which may in turn narrow down the possible locations for the fault in the program. Referring back to the previous example, the fact that the faulty block $C_3$ is executed subsequent to $C_1$ and prior to $C_4$ is irrelevant for the diagnosis.

## 2.3   Related Work

Although to our knowledge the techniques described in this paper are not common practice in the development of consumer electronics embedded software, they have successfully been applied in other systems. However, many of these techniques are usually very expensive in terms of computation and memory space usage and thus they will not work in the embedded software domain due to their peculiarities.

Pinpoint [4] is a framework for root cause analysis on the J2EE platform. Pinpoint can be explained as trace comparison combined with a specific form of error detection. Pinpoint is targeted at large, dynamic internet services.

Reference [5] describes a method for localization of a failure-causing class by comparing method call sequence of passing and failing runs in an object oriented language. A difference in method call sequences is likely to point to the erroneous class.

A comparable technique is described in [6]. It describes a software layer for encoding program execution, i.e., compacting trace data. It is to be complemented with a (bottom) layer for extracting trace data, and a (top) layer for analyzing the compacted data. Trace data is generated at function/method call resolution. Many of the proposed compacting techniques are probably too expensive for real-time systems.

Tarantula [7, 8] is a visualization system that displays the results of running suites of tests. Tarantula logs all executed statements and then compares the traces from pass and failure runs. Executed statements that are present in a failure run and not in a pass run are likely to be the reason of the error. The granularity of the traces makes this technique impossible in the embedded systems domain.

When memory space is a constraint, abstraction of traces is desirable. In *path profiling* [9], a program is instrumented so that the number of times different paths of the program execute is accumulated during an execution. An execution overhead in the range of 30–40% is reported, so path profiling is less attractive for real-time systems. Interprocedural path profiling, which should be even more expensive, is described in [10].

*Path spectra* are one way to compare program behavior, and in [3], they are compared with several other kinds of program spectra, on the Siemens test set of [11]. The comparison is only for regression testing, though: given a program $P$ and a modified version $P'$: if the outputs $P(i)$ and $P'(i)$ of both versions on input $i$ differ, compare spectra $s(P, i)$ and $s(P', i)$ to locate the cause.

Because full tracing seems unfeasible in embedded systems domain mainly because of memory space limitations, we choose to conduct our first experiment in program spectra. As discussed, this technique can be seen as an abstraction of traces. The rest of the paper presents our experimentations with program spectra.

## 3   Code   Instrumentation   Using   FRONT

The tool used to instrument the application with our sensors code is FRONT. FRONT [12] stands for **F**ast **R**endering **O**f **N**on-terminal and **T**ypes and started as a front-end generator for Philips attribute grammars called Elegant [13]. FRONT can generate a full compiler front-end (performing scanning, parsing, abstract syntax-tree construction and symbol-table handling) in Bison, Flex and C. FRONT is written in its own language and in Elegant. It has been used for internal use within Philips for about three years. During this period, many compilers have been built with this tool. FRONT language is a mixture between a formalism for specifying a context-free language in Extended Backus-Naur form (EBNF) style and a notation for type declarations. Hence, FRONT can be seen as an extension to EBNF in order to add information about the data struct to represent the productions. The FRONT compiler generates a set of C type definitions and a Yacc file that maps an input string ac-cepted by the grammar specified onto a data structure that represents the structure of the input string.

Program transformation (code instrumentation) in FRONT can be done using traversal functions generated by the compiler.

We decide to use FRONT because:

- the ANSI C grammar is public available (supporting pre-processed ANSI C source code);

- it automatically generates the data structure (abstract syntax tree) to represent the source code;

- it automatically generates the symbol table (a big advantage if compared with other compiler-generators);

- it also generates the traversal functions.

## Example

We present a very small example to show how to use FRONT to automatically transform C source code. Suppose we would like to add a log function before each function call in our source code in order to generate a trace. We define the following functions:

- `Create_CallExpr(f_name, arg)` - it creates and returns a call expression with name `f_name` and argument `arg`;

- `comma_add_expr(expr1, expr2)` - this function replaces expr1 in the AST by the comma expression `expr1, expr2`.

As was mentioned, FRONT has support for traversals of the automatically generated AST data structure. The function generated by FRONT to traverse a given type *S* related to grammar *X* is:

```
void Traverse_void_S (
        S x, void_X_TraverseTable t);
```

The first argument is the data type to be traversed and the second is a traverse table. The latter argument is basically a list (structure) of arrays of function pointers. For each production *P* in the grammar one array is added in the structure. The first position of the array contains the function to apply before the traversal of the sub-nodes from the productions and the second position to trigger the function after the traversal of the sub-nodes of the current data type. During the traversal of the AST, these arrays are used to decide which. In our example, we need to set the variable which will trigger the execution of a function when `CallExpr` is found because we want to replace all function calls with the function call plus a log function.

```
void logfunc (expression me) {
    expression logexpr;
    logexpr = Create_CallExpr(
        "log", CallExpr_func(me));
    comma_add_expr(me, logexpr);
}
```

```
struct s_void_TraverseTable add_log;
add_log.action_CallExpr[PRE_ACTION] =
                    &logfunc;
```

Now that we have defined the function to trigger based on the `CallExpr` pattern we still need to invoke the function to start the traversal. Actually this traversal can be seen as a source-to-source transformation and is called as follows:

```
Traverse_void_cfront (cfront_root,
                    &add_log);
```

In our experiments we use FRONT to automatically instrument the TV source code in order to obtain traces - in fact in order to compact these traces we generate abstraction of traces as it is explained in Section 2.2.

# 4    Experiment with TV Control Software

In this section we demonstrate that in a typical consumer electronics product, software fault diagnosis based on the comparison of program spectra is feasible.

## 4.1    Test Case

The subject of our experiment is the control software in a particular product line of analog television sets. Since we study an analog set, the essential functionality is implemented in hardware, but everything is coordinated in software, which is responsible for tasks such as decoding signals from the remote control, displaying and navigating through the on-screen menu, and coordinating the audio and video hardware (e.g., setting volume, brightness, and contrast levels). Most teletext functionality, including navigation, and the rendering and caching of pages, is also implemented in software.

The control processor is a MIPS running a small multi-tasking operating system. Essentially, the run-time environment consists of several threads with increasing priorities, that execute a message dispatch loop each. Messages are read from a queue per thread, and trigger the execution of specific functions, to which they contain the arguments. Threads are preempted when a message arrives on a queue associated with a higher-priority thread. In the specific version of the software that we studied, there are just over 300 functions tied to the message dispatch loops.

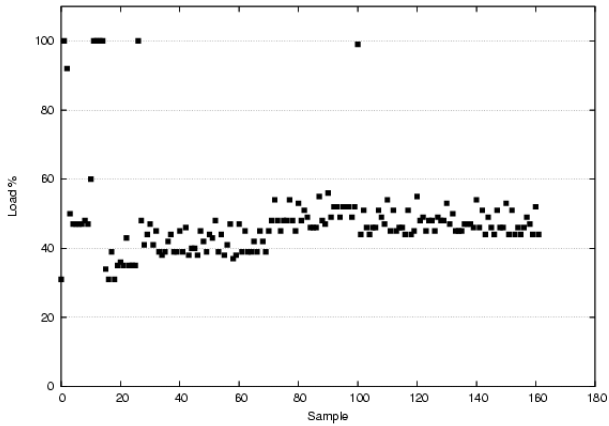

**Figure 2. CPU load measured per second**

All threads share the same address space. The total available memory in consumer sets is two megabytes, but in the special developer set that we used for our experiments, another two megabyte was available. In addition, the developer sets have a serial connection that can be accessed by the control processor, and a debugger interface for manual debugging on a connected PC.

## 4.2 Error Description

A known problem during development time with the specific version of the control software that we had access to, is that after teletext viewing, the CPU load when watching television (TV mode) is approximately 10% higher than before teletext viewing. Figure 2 shows the CPU load measured per second, for the following scenario: one minute TV mode, 15 sec. teletext viewing, 15 sec. transparent teletext viewing, and one minute of TV mode. The CPU load clearly increases around the 70th sample, when the 15 sec. of teletext viewing starts,[2] and never returns to the lower load after sample 100, when we switch back to TV mode.

## 4.3 Recording Program Spectra

To diagnose the load problem, we collected function call spectra for the 300+ functions that are called through the message dispatch loops, described at the beginning of this section. In this case, a program spectrum is an array of just over 300 counters, that shows how often each of the individual functions has been called. For collecting the spectra, we had to deal with the following two issues, which we believe are characteristic for embedded software in consumer electronics.

- No clear concept of a "run" exists for interactive systems such as a television set.

[2]The first 10 samples are a start-up phase, and should be ignored.

- The control software is part of a time-critical system. Especially for timing-related faults, we want to minimize the impact of our measurements on the timing behavior of the system.

Instead of collecting program spectra for runs of the software, for this experiment we collected them periodically, starting with a new set of function call counters every second. In general, it should be possible to identify some notion of a transaction that program spectra can be collected for. A notion of a transaction that we could have used as an alternative for collecting program spectra periodically is the period in between two key presses on the remote control.

Recording a function call spectrum itself has little impact on the timing behavior: in this case, we added one function call to the message dispatch mechanism. This function is called just before handling a message, with the unique integer ID of the handling function as an argument. The function itself uses this ID as an index in the array of counters, and increments the identified counter. For this experiment, we stopped counting at 127.

Extracting the recorded spectra from the television set for off-line diagnosis is more difficult, however. In this case, the communication channel is a serial connection to a PC running a terminal emulator, and we want to postpone sending a recorded spectrum until the CPU becomes idle. For this purpose, we introduced some infrastructure for caching profiles of past transactions in the extra memory that we had available in the development sets. This infrastructure consists of a queue of free spectra, and a queue of spectra that are waiting to be sent to over the serial connection. Every second, the "current" spectrum is added to the latter queue, and a spectrum from the former queue is made current for new measurements.

To minimize the impact on the timing behavior, the actual sending of the spectra is implemented asynchronously via the message dispatch loop on a low priority thread, which only becomes active if the system is otherwise idle. Spectra are communicated one counter at a time, allowing new messages to be dispatched also on this low priority thread in between. Although for this particular experiment, there was enough memory available for storing all spectra in the television set during the execution of the test scenario, this is not possible in general, and we believe that the asynchronous caching and sending of program spectra is essential for applying this technique in consumer electronics embedded software.

## 4.4 Diagnosis

To diagnose the load problem, we compared two sets of ten spectra, one drawn from the first 60 sec. of TV mode, and the other drawn from the second 60 sec. of TV mode in the test scenario. This comparison identified two functions that were called via the message dispatch mechanism only after teletext viewing. One of these functions had already been identified by the developers as the actual location of the fault. So using program spectra, we were able to diagnose the load problem with an accuracy of 50%, yet, by fully automatic means.

# 5    Conclusions and Further Research

From the experiment described in Section 4 we can draw the following conclusions.

- Based on the case study, collecting and analyzing transaction spectra through source code instrumentation seem feasible and useful techniques for obtaining insight in the dynamic behavior of embedded software in consumer electronics devices;

- It has successfully been applied to diagnose a known CPU load problem in a realistic, industrial test case;

Although quite simple in itself, our diagnosis of the CPU load problem can be explained as a form of automatic debugging: by analyzing the differences between spectra of transactions where the problem manifests itself, and spectra of transactions where the systems behaves correctly, we have identified the cause of the problem. In the experiment described in Section 4 we just measure the execution of functions that are called through the message dispatch loop. Following this tactic we can only detect and diagnose errors that are activated by the message dispatch loop. For example, the message loop is sending messages when it should be idle. Obviously if we increase the granularity of the measurements we will get better results. The research question is what is the level of granularity which we can trace without interfering with the behavior of the device. In embedded systems we can not trace everything so we need to investigate another granularity of traces and also other methods of compacting traces. For example we think that basic-blocks spectra will give us insight about errors that are control flow dependent. Thus finding the best way of tracing/profiling basic blocks in the embedded systems domain is our next step.

Fault diagnosis techniques are inherently related to error detection. In turn, the outcome of the diagnosis can be used for recovery of the systems. Manually fixing a software fault that has been identified through automatic debugging is a form of recovery, but we believe that the diagnosis techniques discussed here have applicability beyond the software development phase, and can serve as the basis for automatic recovery strategies. The Pinpoint framework, discussed in Section 2.3, illustrates the role that automatic debugging can play in recovery: Pinpoint identifies components of a J2EE system that are active primarily in transactions that display a significant amount of fault symptoms.

In the case of the large-scale on-line transactional systems for which Pinpoint was developed, these components are typically redundant hardware (servers), that can be reset, or otherwise serviced to recover the system.

In the context of consumer electronics, a possible recovery mechanism could be to restart processes or threads that demonstrate suspicious behavior, or to reset the data structures involved. In the presence of false positives this may lead to unnecessary recovery attempts, but if these go unnoticed, while they prevent more severe recovery measures, such as a system reboot after a full system lock-up, they may well be preferable. In this context, we would like to remark that preventive resetting of processes is an established method to achieve robustness with respect to errors resulting from process aging, which typically involves the depletion of a scarce resource, such as memory. In software rejuvenation [14], this is done periodically, but with error detection and diagnosis in place we may be able to make a better educated guess as to what parts of a system need to be reset, and when they need to be reset. For actual errors instead of false positives this implies a move from preventive measures to reactive measures, but restarting a low-level process immediately after entering a faulty state may yet prevent the error from propagating to a state where more severe recovery measures are required.

# Acknowledgments

We gratefully acknowledge the feedback from the discussions with our TRADER project partners from Philips Research, Philips Semiconductors, Philips TASS, Philips Consumer Electronics, Design Technology Institute, Embedded Systems Institute, IMEC, Leiden University and Twente University.

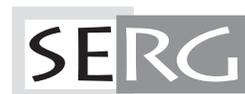